\documentclass[twocolumn,showpacs,amsmath,aps]{revtex4}
\usepackage{amssymb,mathrsfs,bm,color,times}
\usepackage{graphicx,color}
\usepackage{bm}
\usepackage[hypertex]{hyperref}
\usepackage{amsmath}

\newcommand{\be}{\begin{equation}}
\newcommand{\ee}{\end{equation}}
\newcommand{\bea}{\begin{eqnarray}}
\newcommand{\eea}{\end{eqnarray}}
\newcommand{\bsube}{\begin{subequations}}
\newcommand{\esube}{\end{subequations}}
\newcommand{\Fig}[1]{Fig.\,\ref{#1}}
\newcommand{\Eq}[1]{Eq.\,(\ref{#1})}
\newcommand{\Eqs}[1]{Eqs.\,(\ref{#1})}

\newcommand{\la}{\langle}
\newcommand{\ra}{\rangle}

\newcommand{\beq}{\begin{equation}}
\newcommand{\eeq}{\end{equation}}
\newcommand{\beqn}{\begin{eqnarray}}
\newcommand{\eeqn}{\end{eqnarray}}

\newcommand{\nl}{\nonumber \\}

\newcommand{\bsub}{\begin{subequations}}
\newcommand{\esub}{\end{subequations}}

\begin{document}

\title{Cross correlation mediated by Majorana island with finite charging energy}
\author{Wei Feng}
\affiliation{Center for Joint Quantum Studies and Department of Physics,
School of Science, Tianjin University, Tianjin 300072, China}

\author{Lupei Qin}
\affiliation{Center for Joint Quantum Studies and Department of Physics,
School of Science, Tianjin University, Tianjin 300072, China}

\author{Xin-Qi Li}
\email{xinqi.li@tju.edu.cn}
\affiliation{Center for Joint Quantum Studies and Department of Physics,
School of Science, Tianjin University, Tianjin 300072, China}

\date{\today}
\begin{abstract}
Based on the many-particle-number-state treatment for transport
through a pair of Majorana zero modes (MZMs)
which are coupled to the leads via two quantum dots,
we identify that the reason for zero cross correlation of currents
at uncoupling limit between the MZMs
is from a degeneracy of the teleportation and the Andreev process channels.
We then propose a scheme to eliminate the degeneracy
by introducing finite charging energy on the Majorana island
which allows for coexistence of the two channels.
We find nonzero cross correlation established
even in the Majorana uncoupling limit
(and also in the small charging energy limit),
which demonstrates well the teleportation or nonlocal nature of the MZMs.
More specifically, the characteristic structure of coherent peaks
in the power spectrum of the cross correlation is analyzed to identify
the nonlocal and coherent coupling mechanism between the MZMs and the quantum dots.
We also display the behaviors of peak shift
with variation of the Majorana coupling energy,
which can be realized by modulating parameters such as the magnetic field.
\end{abstract}

\maketitle

\section{Introduction}

Majorana zero modes (MZMs) have been widely believed as promising building blocks
for topological quantum computation \cite{Kita01,Kit03,Sar08,Sar15}.
Of the numerous physical realizations of MZMs, the scheme based on
hybrid semiconductor-superconductor devices seems to be
the most advanced leading candidate \cite{Ali12,Flen12,Bee13,Agu17a,Ore18,Kou19,Kou20}.
For identification of the MZMs,
a variety of transport predictions have been proposed
but so far the major evidence is the zero-bias-peak of conductance
\cite{Kou12,XHQ12,XHQ13,Fin13,Mar16,Sar21,Mar17},
while demonstration of the peak height of quantized conductance $2e^2/h$
is still waiting for further efforts \cite{Law09,BNK11,Sarma01,Flen10,Flen16,Zhang21,Zhang21-2}.


We notice that either the zero-bias-peak or the quantized conductance
is based on the setup of single-lead local measurement,
whereas the local measurement is insufficient
to fully confirm the existence of the MZMs.
Despite that the most convincing signals should be the non-Abelian statistics,
the earlier further step can be the measurements
based on nonlocal transport of two-lead devices,
such as the measurement of nonlocal conductance matrix \cite{BCS-1,BCS-2,Nay21},
and in particular the measurement of nonlocal cross correlation of currents
\cite{Dem07,Bee08,Zoch13,Li12,Law09,BNK11,Shen12,Law13,Has15,Ore15,Dev17,Has20,CPB21}.

In a two-lead transport setup, the nonlocal nature of the MZMs implies
both the so-called ``teleportation" and crossed Andreev reflection (CAR) processes.
The former process mediates direct transfer of the same type of particles (electron or hole)
between the two leads, while the latter converts
an incoming electron into an outgoing hole in a different lead.
The both channels are expected to establish cross correlation of currents,
through the MZMs over a distance much longer than
the superconducting coherence length (i.e., the size of Cooper pairs).
Nevertheless, it has been shown that,
in the setup of the MZMs coupled to the leads directly,
nonlocal cross correlation of currents does not exist
at the limit of the Majorana coupling energy
$\epsilon_M\to 0$ \cite{Dem07,Bee08,Zoch13}.
Existing studies of the cross correlation have focused thus on
the case of $\epsilon_M\neq 0$, i.e., the MZMs with finite coupling energy
which mediate a cross correlation with certain unique features \cite{Bee08,Zoch13}.

On the issue of Majorana nonlocality,
the most interesting problem is yet how to establish
nonzero cross correlation at the long distance limit with $\epsilon_M\to 0$.
In Refs.\ \cite{Shen12,Zoch13}, the alternative setup has been considered
by coupling a pair of MZMs to leads via single-level quantum dots (QDs)
as schematically shown in Fig.\ 1.
In this setup, the single-level QDs are expected to suppress
the local Andreev reflection (LAR) process,
thus allowing establishing nonzero cross correlation
through the teleportation and the CAR channels.
However, in Ref.\ \cite{Zoch13},
it was found that the cross correlation is proportional to $\epsilon^2_M$,
implying thus a completely vanished cross correlation at the limit $\epsilon_M\to 0$.
On the contrary, by focusing on the long wire limit
of $\epsilon_M\to 0$ in Ref.\ \cite{Shen12},
nonzero cross correlation was claimed
together with unique features via modulating the QD levels.
The disagreement between Refs.\ \cite{Zoch13} and \cite{Shen12}
was discussed in the Supplementary Material of Ref.\ \cite{Zoch13},
by attributing the reason to the use of `diagonalized' master equation
which may lack certain quantum coherence,
while the latter Erratum of Ref.\ \cite{Shen12}
briefly stated that the reason was from sign errors.

\begin{figure}[ht]
  \centering
  \includegraphics[width=8.0cm,height=2.5cm]{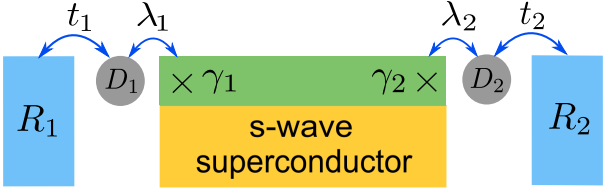}
  \caption{
Schematic setup for transport through a pair of Majorana zero modes (MZMs)
which are coupled to the leads ($R_1$ and $R_2$) via two quantum dots ($D_1$ and $D_2$).
The MZMs are assumed by the semiconductor nanowire realization,
by means of making the nanowire in proximity contact with an s-wave superconductor.
In this work, a finite charging energy on the Majorana island
will be introduced to eliminate the degeneracy of the teleportation
and the Andreev process channels,
resulting thus in nonzero cross correlation of currents
even at the uncoupling limit between the Majorana modes
$\gamma_1$ and $\gamma_2$.    }\label{scheme}%
\end{figure}

Indeed, after correcting the sign errors, the master equation approach
can arrive at the same conclusion of zero cross correlation
at the limit of $\epsilon_M\to 0$, as concluded by
the scattering matrix based Bogoliubov-de Gennes (BdG) treatment in Ref.\ \cite{Zoch13}.
In this work, we clarify that the underlying ``mechanisms"
leading to the vanished cross correlation,
implied by the two solving approaches, are quite different.
In the BdG treatment, the vanished cross correlation is caused by
a cancellation owing to quantum destructive interference
between the positive and negative energy states, or, equivalently,
from uncoupling between the Majorana modes at the limit $\epsilon_M\to 0$.
However, based on a careful examination,
we find that the absence of cross correlation,
in the {\it BdG-free} occupation-number-state treatment,
is owing to a ``degeneracy" of the Andreev process and the teleportation channels.
Inspired by this observation, we propose a scheme to eliminate the `degeneracy'
by introducing finite charging energy $E_C$ on the Majorana island,
which allows subsequently a restoration of nonzero cross correlation
even at the limit $\epsilon_M\to 0$.
This consideration is a generalization of Ref.\ \cite{Fu10},
where a large $E_C$ limit was proposed to fully suppress the Andreev process
and to single out the pure teleportation channel alone for transport.
The rich physics of coherent dynamics
resulting from the interplay of the two channels
can help to confirm the MZMs beyond the zero-bias-peak of conductance.
The present study is also expected to stimulate further study of Majorana island
in the presence of finite $E_C$, which has been commonly assumed in contexts
such as Majorana qubit construction, manipulation, and measurement \cite{Opp10,Egg14,Egg16,Fle16,Ali18,Mar18,Sch20,Mar21,Mar21-2}.

\section{Interplay of Teleportation Transfer and Andreev Process}

Before carrying out a full calculation in Sec.\ III for the cross correlation of currents
in the two-lead (three-terminal) transport setup as schematically show in Fig.\ 1,
where a pair of MZMs are coupled to the leads through two single-level quantum dots (QDs),
in this section we consider the charge transfer dynamics
in the central device of the isolated QD-MZMs-QD system.
This simple dynamics has been studied in Ref.\ \cite{Lee08},
but here our revisit will allow us to gain a key insight for our next step of full study.

\subsection{Insight from charge transfer dynamics in the isolated Dot-MZMs-Dot system}

For the isolated QD-MZMs-QD system, we split the low-energy effective Hamiltonian
into two parts, $H_S = H_0+H'$, with the individual parts given by
\begin{eqnarray} \label{Hams}
H_0&=& i \frac{\epsilon_M}{2} \gamma_1\gamma_2 +\sum_{j=1,2}
\epsilon_j d_{j}^{\dagger} d_{j} \,, \nl
H'&=& \sum_{j=1,2}  \lambda_j  ( d_j^\dagger-d_j)\gamma_j  \,.
\end{eqnarray}
In $H_0$, $\epsilon_M$ is the coupling energy of the two Majorana modes
$\gamma_1$ and $\gamma_2$,
and $\epsilon_j$ is the energy level of the $j_{\rm th}$ QD.
In the tunnel coupling Hamiltonian $H'$, $\lambda_1$ and $\lambda_2$
are the coupling amplitudes of the MZMs to the QDs.
In order to carry out a study for the charge transfer dynamics
in the occupation-number-state representation,
we convert the Majorana modes into regular fermion,
through the transformation $\gamma_1=i(f-f^\dagger)$ and $\gamma_2=f+f^\dagger$, yielding
\begin{eqnarray} \label{Ham1}
H_0&=&\epsilon_M (f^{\dagger} f -\frac12) + \sum_{j=1,2}  \epsilon_j d_{j}^{\dagger} d_{j} \,, \nl
H'&=& \sum_{j=1,2} \lambda_j  \left[ d_j ^{\dagger} f
 + (-1)^{j} d_j ^\dagger f^\dagger +\text {H.c.} \right] \,.
\end{eqnarray}
From the tunnel-coupling Hamiltonian in this form,
it is clear that there exist two processes.
One is the normal tunneling process described by $d_j ^{\dagger} f$,
and its Hermitian conjugate.
The charge transfer path through the $f$ particle,
owing to its highly nonlocal nature,
is usually termed as ``teleportation" channel.
Another process in $H'$, described by $d_j ^\dagger f^\dagger$ and its Hermitian conjugate,
is the famous Andreev process associated with splitting and formation of Cooper pairs.

In terms of the occupation-number-state representation $|n_1 n_f n_2\ra$,
totally we have eight basis states, i.e., with the electron numbers
of the left dot, central MZMs, and the right dot
($n_1$, $n_f$ and $n_2$) being ``0" or ``1", respectively.
To highlight the most challenging issue of nonlocality,
let us consider the case of $\epsilon_M=0$.
Starting with the state $|100\ra$, the quantum evolution will be restricted within
the odd-parity subspace expanded by $\{ |100\ra, |010\ra, |001\ra, |111\ra  \}$.
It is straightforward to obtain the occupation probabilities
of the left and right dots, respectively, as
\begin{eqnarray} \label{P-odd}
P_1^o(t)&=&P_{100}(t)+P_{111}(t)\nl
&=&1-\frac{4\lambda_1^2}{\epsilon_1^2+4\lambda_1^2} \sin^2 \left( \frac{\sqrt{\epsilon_1^2+4\lambda_1^2}}{2}t \right) \,, \nl
P_{2}^o(t)&=&P_{001}(t)+P_{111}(t)\nl
&=&\frac{4\lambda_2^2}{\epsilon_2^2+4\lambda_2^2} \sin^2
\left( \frac{\sqrt{\epsilon_2^2+4\lambda_2^2}}{2}t \right) \,.
\end{eqnarray}
Similarly, starting with the state $|000\ra$, the quantum evolution
is restricted within the even-parity subspace
expanded by $\{ |000\ra, |011\ra, |110\ra, |101\ra  \}$
and yields the occupation probabilities
\begin{eqnarray} \label{P-even}
P_1^e(t)&=&P_{110}(t)+P_{101}(t)\nl
&=&\frac{4\lambda_1^2}{\epsilon_1^2+4\lambda_1^2} \sin^2
\left(\frac{\sqrt{\epsilon_1^2+4\lambda_1^2}}{2}t \right) \,, \nl
P_2^e(t)&=&P_{011}(t)+P_{101}(t)\nl
&=&\frac{4\lambda_2^2}{\epsilon_2^2+4\lambda_2^2} \sin^2
\left(\frac{\sqrt{\epsilon_2^2+4\lambda_2^2}}{2}t \right) \,.
\end{eqnarray}

From the results of \Eqs{P-odd} and (\ref{P-even}),
we find that the occupation probability of the right (left) dot
does not depend on the occupation of the left (right) dot
and the coupling strength $\lambda_1$ ($\lambda_2$).
To be more specific, for instance, the dynamics with the initial state $|100\ra$
involves a transition from the intermediate state $|010\ra$
to $|001\ra$ via the teleportation channel,
and as well a transition from the initial state $|100\ra$ to $|111\ra$
via the local Andreev process on the right side.
Both processes lead to occupation of the right dot with identical effect,
which is also identical to the result of state transition of $|n_f n_2\ra$
between $|10\ra$ and $|01\ra$, or, between $|00\ra$ and $|11\ra$,
by setting $\lambda_1=0$ which cuts the coupling to the left dot completely.
Similarly, the dynamics with the initial state $|000\ra$
involves a transition from $|110\ra$ to $|101\ra$
via the intermediate excitation of the $f$ quasiparticle,
and as well a transition from the initial state $|000\ra$ to $|011\ra$
via the local Andreev process on the right side.
And, being precisely the same,
both channels make the occupation of the right dot indistinguishable,
and identical to the result by setting $\lambda_1=0$.
The above observation implies an important result that,
by coupling the MZMs to transport leads via the QDs,
the currents in the left and right leads should have
no cross correlation at the limit $\epsilon_M\to 0$,
since the disturbance of current measurement on one side
will not affect the current on the other side,
because of the reason explained above.

\subsection{Resolving a discrepancy in literature}

In Refs.\ \cite{Zoch13} and \cite{Shen12}, the same setup as shown in Fig.\ 1
was studied and different conclusions were achieved.
It was concluded in Ref.\ \cite{Zoch13},
based on the single-particle BdG treatment in terms of scattering matrix approach,
that the cross correlation is proportional to $\epsilon^2_M$,
implying thus a completely vanished cross correlation at the limit $\epsilon_M\to 0$.
However, in Ref.\ \cite{Shen12},
applying a many-particle-number-state treatment of master equation approach,
nonzero cross correlation at the limit of $\epsilon_M\to 0$ was claimed
together with unique behaviors via modulating the QD levels.
The disagreement was discussed
in the Supplementary Material of Ref.\ \cite{Zoch13}
by attributing to the use of ``diagonalized" master equation
which may lack certain quantum coherence,
while the later Erratum of Ref.\ \cite{Shen12}
briefly stated that the true reason was from sign errors.
After correcting the sign errors, it was found that the master equation approach
can also give the result of zero cross correlation at the limit of $\epsilon_M\to 0$.
This is in full agreement with the expectation mentioned above from analyzing
the charge transfer dynamics in the isolated Dot-MZMs-Dot system.
Nevertheless, the reason leading to the zero cross correlation
in the BdG treatment \cite{Zoch13}
is quite different from that revealed by
the above occupation-number-state dynamics analysis.

The BdG matrix of Hamiltonian of the central QD-MZMs-QD segment
in Ref.\ \cite{Zoch13} is summarized here as follows
\begin{eqnarray}\label{sp-BdG-1}
H =  \left(
\begin{array}{cccccc}
 0 & i\epsilon_M & \lambda_1 & 0 & -\lambda^*_1  & 0  \\
-i\epsilon_M & 0 & 0 & \lambda_2 & 0 & -\lambda^*_2 \\
\lambda^*_1 & 0 & \epsilon_1 & 0 & 0 & 0 \\
0 & \lambda^*_2 & 0 & \epsilon_2 & 0 & 0 \\
-\lambda_1 & 0 & 0 & 0 & -\epsilon_1 & 0  \\
0 & -\lambda_2 & 0 & 0 & 0 & -\epsilon_2 \\
\end{array}
\right) \,.
\end{eqnarray}
In terms of the BdG single-particle-state picture,
the underlying state basis reads as
$\{ |\Phi_1\ra, |\Phi_2\ra, |e_1\ra, |e_2\ra, |h_1\ra, |h_2\ra \}$,
with $|e_{1,2}\ra$ ($|h_{1,2}\ra$) the electron (hole) states of the quantum dots,
while $|\Phi_1\ra$ and $|\Phi_2\ra$ the Majorana modes (wavefunctions)
associated with the Majorana operators $\gamma_1$ and $\gamma_2$.
It is clear by this formulation that no cross correlation can be established
at the limit $\epsilon_M \to 0$,
since the connection between the two quantum dots cannot be established,
owing to decoupling of the Majorana modes $|\Phi_1\ra$ and $|\Phi_2\ra$
when $\epsilon_M\to 0$.
Alternatively,
one can also use the eigenstates $|E_0\ra$ and $|-E_0\ra$
for the pair of Majorana modes (with $E_0=\epsilon_M$).
In this representation, a cancellation owing to quantum destructive interference
between the positive and negative energy states will also result in the conclusion
that no coupling between the quantum dots can be established when $\epsilon_M\to 0$.

In Ref.\ \cite{Shen12}, as mentioned at the beginning of this subsection,
a many-particle-number-state treatment of master equation approach
was applied to the same problem.
For the central QD-MZMs-QD segment, the many-particle states
are denoted as $|n_1 n_f n_2\ra$, with each particle number equal to ``0" or ``1".
For instance, the odd-parity subspace is expanded by the following basis states
\bea
|1\ra &=&|010\ra   \,, \nl
|2\ra &=& |100\ra   \,, \nl
|3\ra &=&|001\ra   \,, \nl
|4\ra &=& |111\ra   \,.
\eea
Using this order of basis, in Ref.\ \cite{Shen12},
the odd-parity sector of the Hamiltonian matrix was carried out as
\begin{eqnarray}\label{Shen-odd-old}
H^o =  \left(
\begin{array}{cccc}
 0 & \lambda_1 & \lambda_2  & 0  \\
\lambda_1^* & \epsilon_1 & 0 & -\lambda_2 \\
\lambda^*_2 & 0 & \epsilon_2 & \lambda_1 \\
0 & -\lambda^*_2 &\lambda_1^* &\epsilon_1+ \epsilon_2 \\
\end{array}
\right) \,.
\end{eqnarray}
Unfortunately, we notice here an essential sign error
in the matrix elements $H^o_{24}$ and $H^o_{42}$,
owing to ignoring the Fermi-Dirac quantum statistics.
That is, a correct treatment should lead us to the following result
\bea
\lambda_1 d_1^{\dagger}f^{\dagger} |001\ra &=& \lambda_1 |111\ra \,, \nl
\lambda_2 d_2^{\dagger}f^{\dagger} |100\ra &=& -\lambda_2 |111\ra  \,.
\eea
Importantly, the minus sign in the second identity is necessary
owing to the Fermi-Dirac statistics,
caused by the cross order of action
when acting $d_2^{\dagger}f^{\dagger}$ on $|n_f n_2\ra$.
Notice that there is no this type of sign problem
when acting $d_1^{\dagger}f^{\dagger}$ on $|n_1 n_f\ra$.
Therefore, the correct form of the odd-parity sector of
the Hamiltonian matrix should be the following one
\begin{eqnarray}\label{Shen-odd-new}
H^o =  \left(
\begin{array}{cccc}
 0 & \lambda_1 & \lambda_2  & 0  \\
\lambda_1^* & \epsilon_1 & 0 & \lambda_2 \\
\lambda^*_2 & 0 & \epsilon_2 & \lambda_1 \\
0 & \lambda^*_2 &\lambda_1^* &\epsilon_1+ \epsilon_2 \\
\end{array}
\right) \,.
\end{eqnarray}
In the even-parity sector of the Hamiltonian matrix,
similar sign errors exist as well in Ref.\ \cite{Shen12}.

After correcting the sign errors, the master equation approach can give
the same result of zero cross correlation at the limit $\epsilon_M\to 0$.
Thus the ``result-showing discrepancy"
between Refs.\ \cite{Zoch13} and \cite{Shen12} does not exist.
However, the underlying mechanisms
leading to the vanished cross correlation are quite different.
As explained above, the single-particle-state BdG treatment
attributes the reason to no coupling between the left and right quantum dots,
owing to no coupling between the Majorana modes at the limit $\epsilon_M\to 0$.
In contrast, in the many-particle-number-state treatment using $|n_1 n_f n_2\ra$,
it is clear that the occupation of $n_f=1$ (excitation of the $f$ quasiparticle)
can mediate electron transmission between the quantum dots,
owing to the nonlocal nature of the $f$ particle's wavefunction in space,
i.e., the $f$ particle couples simultaneously to the left and right dots
even at the limit $\epsilon_M\to 0$.
As analyzed in detail in Sec.\ II A,
the basic reason leading to zero cross correlation
is the ``degeneracy" of the teleportation and the Andreev process channels,
which makes the charge occupation of each QD independent of one another.
Thus from the total currents
one is unable to distinguish the individual charge transfer channels,
and no cross correlation mediated by the $f$ quasiparticle can be revealed.

\subsection{Further consideration}

The ``degeneracy" observation allows a couple of ways
to reveal the cross correlation at the limit $\epsilon_M\to 0$.
In Ref.\ \cite{CPB21}, in a (two-lead) three-terminal setup,
nonzero cross correlation is calculated
between the branch circuit currents
flowing back from the superconductor to the leads
through the grounding terminal.
This scheme is actually employing the teleportation-channel-supported CAR process
to mediate the cross correlation, after deleting the component current
flowing between the two leads through the teleportation channel.
The work by Fu in Ref.\ \cite{Fu10}
can be regarded as singling out the teleportation channel alone
by introducing a large charging energy $E_C$ to suppress the Andreev process,
which can obviously result in nonzero cross correlation at the limit $\epsilon_M\to 0$.
An interesting consideration to generalize Ref.\ \cite{Fu10} is
to eliminate the degeneracy, but not to completely suppress the Andreev process.
It seems that we have two ways.
One is to differentiate the coupling strengths
of the teleportation channel and the Andreev process to each lead.
Another one is introducing a modest charging energy $E_C$ on the Majorana island
to differentiate the energies associated with the two channels,
owing to different charge occupations.
Noting that the former protocol should be difficult for experimental realization,
we thus consider the latter scheme in this work.
More details of the model consideration are referred to Sec.\ III B.

Here we may briefly mention that in the latter scheme,
in order to realize the gate-modulated charging energy
through the well known Coulomb-blockade model \cite{Naza09},
the Majorana island should be considered floating.
This type of setup can make the island relatively well isolated,
with thus quantized numbers of total electrons on the island.
Obviously, this consideration is fully compatible with
the ``degeneracy" analysis of the isolated QD-MZMs-QD system.
In reality, after fixing the transport bias voltage,
the central island (together with the proximitized superconductor)
will automatically modulate its Fermi surface
and reach a current-conserving steady state of transport.
In theoretical treatment, being equivalent,
one can fix the Fermi surface of the central island
and properly adjust the chemical potentials of the two leads,
in order to guarantee equal steady-state currents in the left and right leads.
For a symmetric coupling device, the bias voltage ($V$) can be simply applied
as $\mu_1/e=-\mu_2/e=V/2$, with $\mu_1$ and $\mu_2$ the chemical potentials
of the left and right leads, while setting the Fermi surface of
the central island at zero energy as usual.

\section{Nonzero Cross Correlation after Degeneracy Elimination}

In this section, based on the insight gained above,
we consider to introduce a modestly finite charging energy to eliminate the degeneracy
and show the nonzero cross correlation of currents at the limit $\epsilon_M\to 0$.
We first outline in Sec.\ III A
the master equation approach and the technique of noise spectrum calculation,
then specify in Sec.\ III B the many-particle states
involving in the charge transport dynamics,
under proper assumptions of bias voltage window and charging energy scales.
Finally, in Sec.\ III C,
we present the numerical results, physical interpretations and discussions.

\subsection{Master equation approach}

The basic idea of master equation approach to quantum transport
is regarding the central device,
in terms of the language of quantum dissipation,
as the system-of-interest, while regarding the transport leads
as environment \cite{Li05,Qin19}.
In this work we consider to couple the central Dot-MZMs-Dot device
to transport leads to form a two-lead (three-terminal) setup
(with the Majorana island grounded).
The coupling is described by the Hamiltonian
\bea
H'= \sum_{j=1,2}\sum_{k} (t_j d^\dagger_j c_{j,k}+ \text{H.c.}) \,,
\eea
where $c_{j,k}$ are the annihilation operators of the $j_{\rm th}$ lead electrons
which are coupled to the quantum dots with amplitudes $t_j$.
In the weak coupling regime,
under the Born-Markov approximation,
the master equation for the reduced density matrix $\rho$
of the central system reads as \cite{Li05,Qin19}
\begin{equation}
\dot{\rho}= \mathcal{L} \rho
=-i[H_S,\rho]-\frac{1}{2} \sum_{j=1,2}
\{[d_j^{\dagger},D_j^{(-)} \rho-\rho D_j^{(+)}] + \text{H.c.} \}  \,.
\end{equation}
In the second part of dissipative terms, we introduced
\begin{eqnarray}
D_j^{(\pm)} = \int^{\infty}_{-\infty} dt C_j^{(\pm)}(t) e^{\pm i H_S t} d_j e^{\mp i H_S t}.
\end{eqnarray}
The correlation functions of the lead-electrons read as
$C_j^{(+)}(t)=\sum_k |t_j|^2 \langle c_{j,k}^{\dagger}(t) c_{j,k}(0)  \rangle $ and
$C_j^{(-)}(t)=\sum_k |t_j|^2 \langle c_{j,k}(t)  c_{j,k}^{\dagger}(0) \rangle$,
where the statistical average is referred to the local-thermal-equilibrium of leads
and the time dependence is from the interaction picture with respect to the leads Hamiltonian.
Under wide-band approximation for the leads, the matrix elements of $D_j^{(\pm)}$
can be straightforwardly carried out using the eigenstate basis
of the central system Hamiltonian $H_S$, as \cite{Li05,Qin19}
\begin{eqnarray}
 \left( D_j^{(\pm)} \right)_{nm}=\Gamma_j f^{\pm}_j(\omega_{mn}) (d_j)_{nm}  \,.
\end{eqnarray}
Here $\omega_{mn}$ is the energy difference of the eigenstates, say, $\omega_{mn}=E_m-E_n$,
while $f^{\pm}_j$ are the Fermi occupied and unoccupied functions of the $j_{\rm th}$ lead.
The tunnel-coupling rates are introduced through $\Gamma_j=2\pi \nu_j |t_j|^2$,
with $\nu_j$ the density-of-states of the $j_{\rm th}$ lead.

In the master equation approach, the transient current in the $j_{\rm th}$ lead
can be conveniently expressed as
$I_j(t)={\rm Tr} [\hat{I}_j \rho(t)]\equiv \la \hat{I}_j\ra$,
while the current operator is given by \cite{Li05,Qin19}
\begin{eqnarray}
\hat{I}_j =\frac{1}{2} \left( d_j^\dagger D_j^{(-)}
- D_j^{(+)}d_j^\dagger \right) + \text{H.c.}  \,.
\end{eqnarray}
Moreover, the master equation approach allows a very convenient way
to calculate the current correlation functions, by taking the spirit
of quantum regression theorem and a generalized quantum-jump technique.
Specifically, for the cross correlation of our interest in this work, we introduce
$S_{RL}(t) =  \la \delta \hat{I}_2 (t) \delta \hat{I}_1 (0) \ra$,
where $ \delta \hat{I}_j(t)= \hat{I}_j(t)- \la \hat{I}_j \ra $.
The key quantity we need to calculate is \cite{Li05,Qin19}
\begin{eqnarray}\label{S-RL}
\widetilde{S}_{RL}(t) = \text{Tr} [ {\hat{I}_2} e^{\mathcal{L}t}
(\hat{I}_1\triangleright\bar{\rho})]  \,.
\end{eqnarray}
Here $\bar{\rho}$ is the steady state of the central device
and we introduced the notation
\bea
\hat{I}_1 \triangleright \bar{\rho} \equiv
\frac{1}{2} \left(  D_1^{(-)}\bar{\rho}d_1^\dagger
- d_1^\dagger \bar{\rho}D_1^{(+)} \right) + \text{H.c.}
\eea
Then, the cross correlation of current fluctuations
is given by $S_{RL}(t)=\widetilde{S}_{RL}(t)-\bar{I}_2 \bar{I}_1$,
where $\bar{I}_2$ and $\bar{I}_1$ are the steady-state currents in the two leads.

The power spectrum $S_{RL}(\omega)$ of the cross correlation of currents
is simply the Fourier transformation of $S_{RL}(t)$.
In practice, $S_{RL}(\omega)$ can be obtained more directly
by Laplace-transforming the master equation for $\widetilde{\rho}(t)$,
which is the evolving state governed by
$\widetilde{\rho}(t)=e^{\mathcal{L}t}\widetilde{\rho}(0)$,
while the ``initial" state after the first current measurement
is given by $\widetilde{\rho}(0)=\hat{I}_1 \triangleright \bar{\rho}$.

\subsection{State basis}

To eliminate the degeneracy of the teleportation and local AR channels,
we introduce charging energy $E_C$ on the Majorana island
which, however, does not suppress the AR process,
being thus a generalization of the work by Fu \cite{Fu10}.
Applying the well-known Coulomb blockade model \cite{Naza09},
we introduce the charging energy term
\bea
H_{C}=E_C(\hat{n}-n_g)^2  \,,
\eea
being added to the Dot-MZMs-Dot system Hamiltonian $H_S$ given by \Eq{Hams}.
The charging energy reads as $E_C=e^2/C$,
with $C$ the effective capacitance of the Majorana island.
$\hat{n}$ is the quantized net-electron-number operator on the island,
and $n_g$ is the gate charge (continuous variable)
which can be adjusted by gate voltage.
Let us denote the eigenvalues of the number operator $\hat{n}$ as $\tilde{n}=n_f+2n_c$,
where $n_f$ is the occupation number of the MZMs associated $f$ particle,
and $n_c$ is the number of Cooper pairs created or destroyed by the Andreev process.
Let us also introduce $\widetilde{E}_C=E_C(\tilde{n}-n_g)^2$,
for the respective charging energies.

In order to carry out the solution of the master equation
and the spectrum of the cross correlation function of currents,
we specify the state basis for the central system as $|n_1 n_f n_2,2n_c\ra$,
with $n_1$ and $n_2$ the electron numbers of the left and the right dots,
and $n_f$ and $n_c$ the numbers of the $f$ quasiparticle
and the created or destroyed Cooper pairs.
The total electron number of the central Dot-MZMs-Dot device
is $N=n_1+n_f+n_2+2n_c$,
which is restricted in our calculation as $N = -1, 0, 1$ and $2$
by setting the gate charge $n_g=1/2$
and the charging energy $\widetilde{E}_C \leq \frac{9}{4}E_C$.
Here we have assumed that the next possible charging energy
$\widetilde{E}_C=\frac{25}{4}E_C$
is excluded by the choice of bias voltage window.
Under these considerations, all the many-particle states
which are involved in the transport dynamics
and in our numerical calculation are listed in Table I.

\begin{table}[h]
\normalsize
\caption{
Basis states involved in the transport dynamics
and the associated charging energies
$\widetilde{E}_C=E_C(\tilde{n}-n_g)^2$,
by setting the gate charge $n_g=1/2$.
The total electron number in the Dot-MZMs-Dot system
is denoted by $N=n_1+n_f+n_2+2n_c$,
while the electron number on the Majorana island
is denoted by $\tilde{n}=n_f+2n_c$.  }
\begin{tabular}{c|l|c}
  \toprule
  $N$  &  \ \ \ ~States~  &  $\widetilde{E}_C/E_C$\\
\hline
  $ $ & ~$|000,0\ra$ ~&$1/4$ \\
  $0$ & ~$|011,-2\ra$ ~&$9/4$ \\
  $ $ & ~$|110,-2\ra$ ~&$9/4$ \\
 \hline
  $ $ & ~$|100,0\ra$ ~&$1/4$\\
  $1$ & ~$|010,0\ra$ ~&$1/4$\\
  $ $ & ~$|001,0\ra$ ~&$1/4$\\
  $ $ & ~$|111,-2\ra$ ~&$9/4$\\
 \hline
  $-1$ & ~$|010,-2\ra$ ~&$9/4$\\
 \hline
  $ $ & ~$|011,0\ra$ ~&$1/4$\\
  $2$ & ~$|101,0\ra$ ~&$1/4$\\
  $ $ & ~$|110,0\ra$ ~&$1/4$\\
  $ $ & ~$|000,2\ra$ ~&$9/4$\\
 \hline
  $ $ & ~$|111,0\ra$ ~&$1/4$\\
  $3$ & ~$|100,2\ra$ ~&$9/4$\\
  $ $ & ~$|001,2\ra$ ~&$9/4$\\
 \hline
  $4$ & ~$|101,2\ra$ ~&$9/4$\\
 \toprule
\end{tabular}\\
\end{table}

\subsection{Results and discussions}

Let us consider first the special case suggested by Fu in Ref.\ \cite{Fu10},
where the large charging energy $E_C$ is proposed to fully suppress
the Cooper pair splitting and formation processes,
thus singling out the teleportation channel alone.
In the occupation-number-state basis,
there are 8 states involved in the transport dynamics.
In particular, there are two groups of states (two subspaces)
which support coherent dynamics
owing to transition among the states in each subspace,
i.e., the subspace expanded by $\{ |100,0\ra, |010,0\ra, |001,0\ra \}$
with eigen-energies $(-0.27, 1.3, 2.57)$,
and the subspace expanded by $\{ |110,0\ra, |101,0\ra, |011,0\ra \}$
with eigen-energies $(0.03, 1.3, 2.87)$.
In this numerical example, the parameters used are given in the caption of Fig.\ 2.
The characteristic peaks in the current power spectrum reflects
coherent oscillations inside the central system,
with characteristic frequencies as the energy differences of the eigenstates.
This is the basic physics that results in the three characteristic frequencies
we observe in Fig.\ 2,
i.e., $\omega_1=1.27$, $\omega_2=1.57$ and $\omega_3=2.84$.
We may mention that, from the eigenstate energies,
the two groups of states share the same characteristic frequencies.
Note also that tunnel coupling of the central system to the leads simply
causes state switching between the two subspaces analyzed above,
and the other two states $|000\ra$ and $|111\ra$.
These transitions are incoherent
and will not result in coherent peaks in the current power spectrum.

\begin{figure}[ht]
  \centering
  \includegraphics[width=7.5cm]{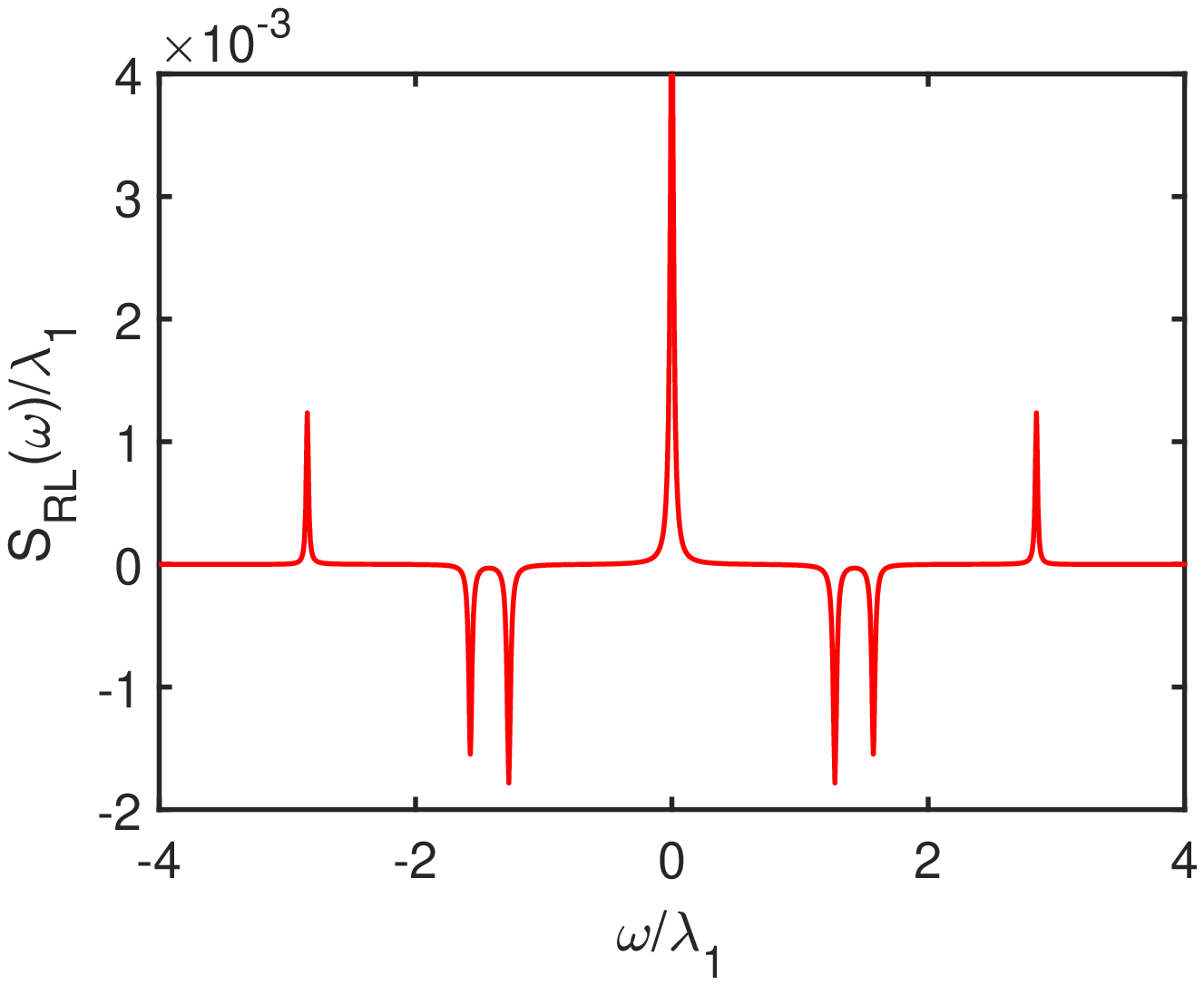}
  \caption{
Cross power spectrum $S_{RL}(\omega)$ for the special case considered in Ref.\ \cite{Fu10},
where a large charging energy $E_C$ on the Majorana island
is assumed to fully suppress the Andreev process.
The cross correlation shown here is thus established by the teleportation channel alone.
In this work (here and in the following), we use an arbitrary system of units
by setting $\lambda_1=\lambda_2=1.0$ and assuming other parameters as:
the bias voltage $\mu_1=-\mu_2=2.0$,
the dot levels $\epsilon_1=\epsilon_2=0.3$,
the Coulomb charging energy $E_C=4.0$,
the coupling rates $\Gamma_1=\Gamma_2=0.02$,
and the Majorana coupling energy $\epsilon_M=0.0$.                                     } \label{largeC}
\end{figure}

The results discussed above under the large $E_C$ limit,
based on the teleportation channel alone,
are relatively simple in physics.
The coherent dynamics is essentially the same as in a triple-quantum-dot system.
A more interesting regime is the intermediate one,
with modestly large charging energy $E_C$
to expose the existence of the teleportation channel,
but at the same time with the Andreev process unsuppressed.
This is a regime to bridge the gap between $E_C=0$ and the large $E_C$ limit,
holding rich physics and results allowing us
to infer the Majorana-mediated quantum dynamics
and the coexistence of different transport channels.

In the presence of charging energy, in contrast to the case of $E_C=0$,
the most important result is that the cross correlation of currents is nonzero.
The physical interpretation is not obvious at all, since in both cases
the teleportation channel and the Andreev process
commonly participate in the transient quantum dynamics.
As analyzed in Sec.\ II, the cross correlation cannot be established
in the ideal Majorana limit ($\epsilon_M=0$), in the absence of charging energy.
The appearance of nonzero cross correlation in the presence of finite $E_C$
differs also from the case of Andreev bound states (ABSs), which may be formed
near the ends of the quantum wire (the central island under present study).
For the case of ABSs, the individual local AR processes
can just maintain independent local currents,
but cannot establish cross correlation
owing to lacking either the crossed AR or the teleportation channel.

\begin{figure}[ht]
  \centering
  \includegraphics[width=7.5cm]{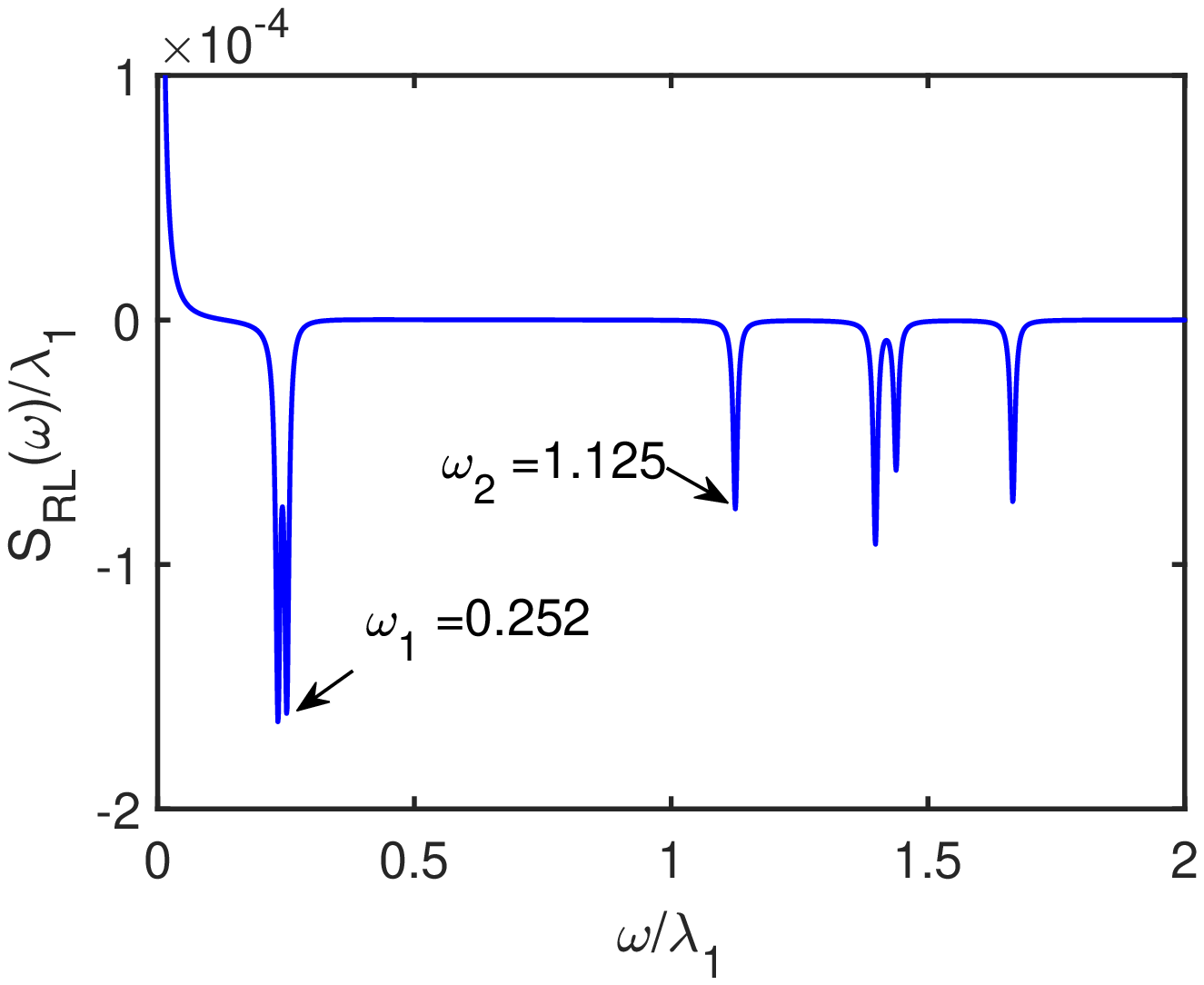}
  \caption{
Cross power spectrum (in the low-frequency regime)
for modestly large charging energy $E_C$
which allows for coexistence of the teleportation and the Andreev process channels.
Using the same arbitrary system of units as in Fig.\ 2
by setting $\lambda_1=\lambda_2=1.0$,
we assume the other parameters in present calculation as:
$E_C=4.0$, $\mu_1=-\mu_2=10.0$, $\Gamma_1=\Gamma_2=0.005$,
$\epsilon_1=\epsilon_2=0.3$, and $\epsilon_M=0.0$.           } \label{finiteC}
\end{figure}

In the presence of $E_C$,
more complicated states participate in the transient dynamics
and result in more complicated spectral structures in the full spectrum.
In Fig.\ 3, we show only the characteristic spectral structures at low frequencies,
which can reflect the essential physical mechanisms
of governing the coherent quantum dynamics.
For instance, the peak at $\omega_1=0.252$ shown in Fig.\ 3
is associated with the Andreev process,
which leads to state transition between $(|100,2\ra, |001,2\ra)$ and $|111,0\ra$.
More quantitatively, with the parameter values given in the caption of Fig.\ 3,
we attribute this characteristic frequency
to the difference of the eigen-energies $E_1=9.3$ and $E_2=9.552$,
with the eigenstates calculated as
\bea
|\Psi_1\ra &=& 0.707|001,2\ra-0.707|100,2\ra \,,   \nl
|\Psi_2\ra &=& 0.696|100,2\ra-0.175|111,0\ra+0.696|001,2\ra \,.  \nonumber
\eea
Note that the quantum supposition of the component states reflects,
from an alternative perspective, the quantum dynamics governed by
the Andreev process of Cooper pair's formation and splitting.

The next peak at $\omega_2=1.125$ shown in Fig.\ 3 is associated with
quantum dynamics governed by both the Andreev process and
the Majorana teleportation channel, which result in state transition
between $(|100,0\ra, |001,0\ra)$ and $(|010,0\ra,|111,-2\ra)$.
More specifically, this characteristic frequency corresponds to
the energy difference of $E_3=1.3$ and $E_4=2.425$,
with the respective eigenstates calculated as
\bea
&& |\Psi_3\ra = 0.707|100,0\ra - 0.707|001,0\ra   \,, \nl
&& |\Psi_4\ra = 0.497|100,0\ra + 0.698|010,0\ra  \nl
&& ~~~~~ + 0.497|001,0\ra + 0.139|111,-2\ra  \,. \nonumber
\eea
Similarly, the quantum superposition of the various component states,
e.g., in $|\Psi_4\ra$, indicates coherent transitions
between the component states, i.e., the transition
$|100,0\ra \leftrightarrow |010,0\ra \leftrightarrow |001,0\ra $
mediated by the Majorana teleportation channel, and the transition
$(|100,0\ra,|001,0\ra)  \leftrightarrow  |111,-2\ra $
owing to the local Andreev process.

Alternatively, in time domain, let us consider a superposition of
$|\Psi_3\ra$ and $|\Psi_4\ra$ for the isolated central system,
e.g., with $|\Psi(0)\ra=(|\Psi_3\ra+|\Psi_4\ra)/\sqrt{2}$ as an initial state.
In Fig.\ 4, we illustrate the time evolution of the state which shows
clear dynamics of quantum oscillations via the transient change
of the occupation probabilities of the component states.
Essentially, it is just this quantum oscillation
that manifests itself as the spectral peak at $\omega_2$ in Fig.\ 3,
in the full system simulation for the cross correlation of currents.

\begin{figure}[ht]
  \centering
  \includegraphics[width=8cm]{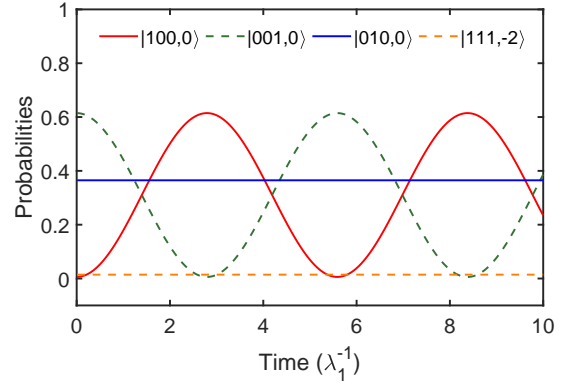}
  \caption{
Interpretation for the characteristic frequency $\omega_2=1.125$ in \Fig{finiteC}.
In time domain, a superposition of $|\Psi_3\ra$ and $|\Psi_4\ra$
for the isolated central system,
e.g., $|\Psi(0)\ra=(|\Psi_3\ra+|\Psi_4\ra)/\sqrt{2}$,
is assumed as an initial state,
and transient probabilities of the individual component states are shown.
The quantum oscillation contributes
the characteristic frequency $\omega_2=1.125$ in \Fig{finiteC}.
The parameters used for this plot are the same as in \Fig{finiteC}.                } \label{Fig4}
\end{figure}

So far, the most important prediction of this work is that
the cross correlation of currents is nonzero even at the limit
$\epsilon_M \to 0$, as shown in Fig.\ 3 and discussed above.
Of course, for the nonideal case of $\epsilon_M\neq 0$,
the cross correlation is nonzero, as already shown in Refs.\ \cite{Bee08,Zoch13}.
However, in the absence of $E_C$, it was predicted that
the cross correlation will vanish with $\epsilon_M\to 0$
according to the $\sim \epsilon^2_M/\Gamma$ behavior \cite{Zoch13},
where $\Gamma$ is the coupling rate to the leads.
Therefore, it is important for experimental verification to demonstrate
the $\epsilon_M$ effect via modulating the system parameters
such as the magnetic field and/or the gate voltage (chemical potential).

 \begin{figure}[ht]
  \centering
  \includegraphics[width=9.5cm]{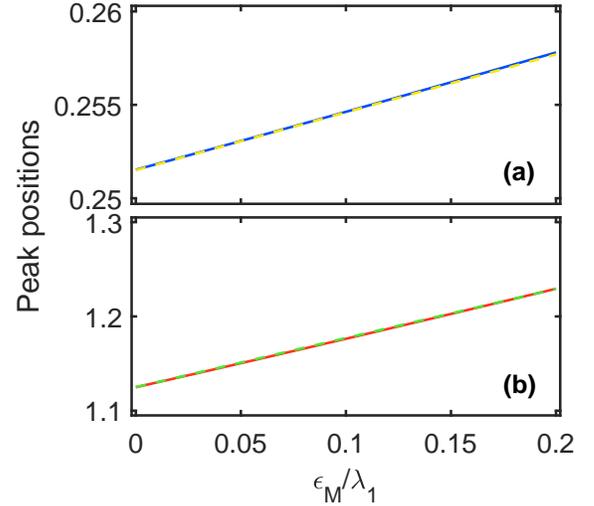}
  \caption{
Peak shift by modulating the Majorana coupling energy $\epsilon_M$.
The peaks at $\omega_1=0.252$ and $\omega_2=1.125$ in \Fig{finiteC}
are exemplified and the results are shown in (a) and (b), respectively.
The solid lines plot the numerical results, while the dashed lines
plot the fitting results of $\omega_1=0.2515+0.031\epsilon_M$
and $\omega_2=1.125+0.521\epsilon_M$.
The parameters used in this plot are the same as in \Fig{finiteC}.    } \label{Fig5}
\end{figure}

The first-step important observation
should be the coherent peaks in the current power spectrum,
since this is possible only in the presence of discrete subgap bound states.
Moreover, in the transport under subgap voltage bias,
the result of nonzero cross correlation
requires that the subgap state must be nonlocal,
which should be simultaneously coupled to the two quantum dots.
Actually, this observation indicates the existence of
a pair of Majorana modes with, most probably, coupling energy $\epsilon_M \neq 0$.
The next step is to reduce $\epsilon_M$ by properly tuning the system parameters.
If the system does take transition from nonideal to ideal phase,
one should observe gradual shift of frequencies
of the coherent peaks, and the coherent peaks
will not vanish when $\epsilon_M \to 0$ as shown in Fig.\ 3.

In Fig.\ 5, we show the peak shift behavior
by modulating the Majorana coupling energy $\epsilon_M$.
For the peak at $\omega_1=0.252$ in Fig.\ 3 under the ideal case $\epsilon_M=0$,
we show in Fig.\ 5(a) the $\epsilon_M$ dependence of this peak,
which can be fitted as $\omega_1=0.252+0.031\epsilon_M$
in the regime of small $\epsilon_M$.
Similarly, for the peak at $\omega_2=1.125$,
we show its $\epsilon_M$ dependence in Fig.\ 5(b),
together with its linear fitting by $\omega_2=1.125+0.521\epsilon_M$.

\section{Small Charging Energy Limit}

In the previous section, we considered (relatively large) finite charging energy $E_C$
which generalizes Ref.\ \cite{Fu10} by allowing the Andreev process unsuppressed.
Importantly, it is constructive to consider small $E_C$ to show
how it can result in cross correlation
by slightly removing the degeneracy of the teleportation and Andreev process channels.
This can support, more evidently, the main conclusion drawn in this work
that the vanished cross correlation in the limit $\epsilon_M\to 0$
is owing to the channels degeneracy, but not the absence of the teleportation channel.

For the convenience of description,
let us denote the state of the Majorana island as $|n_f[n_c]\ra$,
with $n_c$ explicitly accounting for the net numbers
of the Cooper pairs in the island.
Accordingly, we reexpress the states $|n_1 n_f n_2, 2n_c\ra$
(introduced in Sec.\ III) as $|n_1,n_f[n_c],n_2\ra$.
In the main study of this work,
we restrict $n_c=0$ and $\pm 1$,
owing to the relatively large charging energy $E_C$
with respect to the bias voltage.
After including the states of the two quantum dots,
totally we have 24 states of $|n_1,n_f[n_c],n_2\ra$.
However, as shown in Table I (in Sec.\ III B),
we kept only 16 states and removed the other 9 states,
from the restriction of energy conservation.

In the limit of small charging energy $E_C$,
an obvious difficulty in numerical simulation is
that one should include a large number of island states
with $n_c=0, \pm 1, \pm 2, \cdots$.
In the case of $E_C=0$, this difficulty was avoided
by taking into account only the parity states,
i.e., the odd parity state $|n_f=1\ra$ and the even state $|n_f=0\ra$.
This means that we have identified
all the states $|n_f[n_c]\ra$ with different $n_c$
as a single state $|n_f\ra$.

In the presence of charge transfer between the Majorana island and the quantum dots,
the island associated with $n_c=0$ and $\pm 1$
will suffer state transition among 6 states.
For instance, we have
\bea
&|0[-1]\ra \Leftrightarrow |1[-1]\ra \Leftrightarrow |0[0]\ra  \,,   \nl
&|0[0]\ra \Leftrightarrow |1[0]\ra \Leftrightarrow |0[1]\ra   \,.
\eea
In each line of the transitions,
the former is the teleportation channel (normal tunneling process),
while the latter is the Andreev process.
In the limit $E_C\to 0$, the two channels are energetically degenerate.
However, if $E_C\neq 0$, the degeneracy is removed
and cross correlation is expected to establish,
as shown in previous studies of this work.
Further, let us look at the following transitions
\bea
&|1[-1]\ra \Leftrightarrow |0[-1]\ra \Leftrightarrow |1[-2]\ra  \,,   \nl
&|0[1]\ra \Leftrightarrow |1[1]\ra \Leftrightarrow |0[2]\ra   \,.
\eea
If we restrict $n_c=0$ and $\pm 1$, the two states $|1[-2]\ra$ and $|0[2]\ra$
will be absent in the simulation.
Therefore, we cannot properly account for the degeneracy of the teleportation
and Andreev process channels at the limit  $E_C\to 0$,
and thus cannot recover the result of zero cross correlation at this limit.

In order to overcome this difficulty,
let us apply a `boundary condition'
by setting $|0[2]\ra = |0[-1]\ra$ and $|1[-2]\ra = |1[1]\ra $.
This treatment allows us to hold {\it complete} pairs of
the degenerate channels of teleportation and Andreev process,
within the total 24 states associated with $n_c=0$ and $\pm 1$.
We can thus recover the correct result of $E_C=0$.
Based on this treatment, after introducing small nonzero $E_C$,
we can simulate the effect of nonzero cross correlation
established by removing the channels degeneracy, in a continuous manner.

In Fig.\ 6, we display the results of numerical simulations.
Rather than in frequency domain,
we show here the cross correlation function $S_{RL}(t)$ in the time domain,
which can be easily obtained
based on the jump-technique calculation given by \Eq{S-RL}.
The underlying physics is clear:
disturbing the left dot occupation (from its steady state)
by current measurement in the left lead,
the right-lead current will respond in accordance.
This is the nonlocal cross correlation,
which can be nonzero even in the limit $\epsilon_M\to 0$.
However, as indicated in Fig.\ 6,
the cross correlation will vanish asymptotically when $E_C\to 0$.

\begin{figure}[ht]
  \centering
  \includegraphics[width=8.5cm]{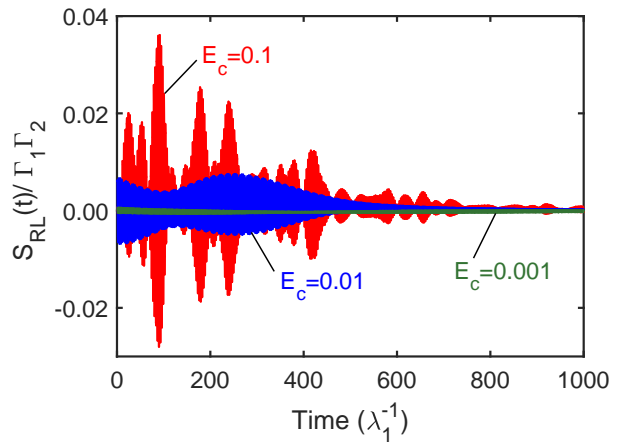}
  \caption{
Illustration for how removing the channel degeneracy
by introducing a small charging energy
can restore nonzero cross correlation
between the currents in the left and right leads.
In this plot, more intuitively,
we show the correlating behavior in time domain
based on the jump-technique calculation given by \Eq{S-RL}.
Results for charging energies of $E_C=0.1$, 0.01 and 0.001 are displayed,
which clearly indicate that the cross correlation will vanish when $E_C\to 0$,
owing to the channel degeneracy discussed in detail in the main text.
Other parameters used here are the same as in Fig.\ 3.    } \label{Fig_Ec}
\end{figure}

\section{Summary and Discussions}

Based on the many-particle-number-state treatment
for transport through a ``Dot-MZMs-Dot" system,
we clarified that the absence of cross correlation
at the Majorana uncoupling limit ($\epsilon_M\to 0$)
is owing to a degeneracy of the Andreev process
and the teleportation channels.
We further proposed a scheme to eliminate the degeneracy
by introducing charging energy $E_C$ on the Majorana island,
which allows a restoration of nonzero cross correlation
even at the limit $\epsilon_M\to 0$.
This result demonstrates well the nonlocal nature of the MZMs.
Specifically, we analyzed the characteristic structure of coherent peaks
in the power spectrum of the cross correlation, which reflect
the nonlocal and coherent coupling mechanism between the MZMs and the quantum dots.
We also displayed the behavior of peak shift
with variation of the Majorana coupling energy,
in a hope for possible demonstration by experiments.

Experimental challenges for studying
the cross correlation analyzed in the present work
should be similar as for previous works \cite{Dem07,Bee08,Zoch13,Shen12}.
However, verification of nonzero cross correlation at the limit $\epsilon_M\to 0$
by gradually introducing a finite (especially small) charging energy
is of great interest,
since it exposes the existence of the Majorana teleportation from the degenerate channels.
It is well known that Majorana teleportation, or non-locality,
is one of the central issues at the heart of Majorana physics.

We noticed that, in the single-particle BdG treatment
in terms of scattering matrix approach,
the vanished cross correlation at the limit $\epsilon_M\to 0$
is caused by a cancellation owing to quantum destructive interference
between the positive and negative energy states, or, equivalently,
from uncoupling between the Majorana modes at the limit $\epsilon_M\to 0$.
This picture looks quite different from the mechanism of {\it degeneracy}
of the two channels revealed by the BdG-free number-state treatment,
where each of the channels supports a non-vanishing charge transfer process.
Searching for a unified understanding of the two pictures
is remained as an open and interesting problem for future investigation.
In the presence of the Coulomb charging energy, it is well known that
the single-particle $S$ matrix scattering approach does not work \cite{Fle16,Mar16N,Sarm18b}.
It seems thus quite challenging to make the many-particle-states treatment
fall into the category of the single-particle BdG scattering approach.
In our opinion, the many-particle-states based channels-degeneracy and its lift
by introducing the charging energy $E_C$
are beyond the scope of the single-particle BdG treatment.

\vspace{0.5cm}
{\flushleft\bf Acknowledgements} \\
This work was supported by the
National Key Research and Development Program of China
(No.\ 2017YFA0303304) and the NNSF of China (Nos.\ 11675016, 11974011 \& 61905174).

\end{document}